\documentclass{article}
\usepackage{framed}

\usepackage{PRIMEarxiv}
\usepackage{siunitx}

\usepackage[english]{babel}
\usepackage[authoryear]{natbib}
\usepackage{amsmath}
\usepackage{mathtools}
\usepackage{comment}
\usepackage{graphicx}
\usepackage[colorlinks=true, allcolors=blue]{hyperref}
\usepackage{standalone}
\usepackage{soul}
\usepackage{ulem}
\usepackage{authblk}
\usepackage{array, makecell}
\usepackage{multirow}
\usepackage{subscript}
\usepackage{xspace}
\usepackage{ifpdf}
\usepackage{ifxetex}
\usepackage{ifluatex}
\usepackage{mathtools}
\usepackage{bm}
\usepackage{graphicx}
\usepackage{tabularx}
\usepackage{textcase}
\usepackage{dashrule}
\usepackage{ragged2e}
\usepackage{authblk}
\usepackage{afterpage}

\newcommand{\figref}[1]{\figurename~\ref{#1}}
\newcommand{\tabref}[1]{\tablename~\ref{#1}}
\renewcommand{\cite}[1]{\citep{#1}}
\newcommand{\TS}[1]{{TS}\textsubscript{#1}\xspace}
\newcommand{\TSj}{\TS{j}}
\newcommand{\Ss}[1]{{SS}\textsubscript{#1}\xspace}

\newcommand{\COtwoe}{CO\textsubscript{2}e}
\DeclarePairedDelimiter{\ceil}{\lceil}{\rceil}

\title{Estimating the carbon footprint of digital agriculture deployment: a parametric bottom-up modelling approach
\thanks{\textit{\underline{Citation}}: 
\textbf{La Rocca et al. 2024. Journal of Industrial Ecology. DOI:10.1111/jiec.13568.}} 
}

\author[1]{Pierre La Rocca}
\author[1]{Gaël Guennebaud}
\author[1,2]{Aurélie~Bugeau}
\author[3]{Anne-Laure~Ligozat}

\affil[1]{Univ. Bordeaux, CNRS, Bordeaux INP, Inria, LaBRI, UMR 5800, F-33400 Talence, France}
\affil[2]{IUF}
\affil[3]{Université Paris-Saclay, CNRS, ENSIIE, LISN, 91400, Orsay, France}

\begin{document}

\maketitle

\begin{abstract}
Digitalization appears as a lever to enhance agriculture sustainability. However, existing works on digital agriculture's own sustainability remain scarce, disregarding the environmental effects of deploying digital devices on a large-scale. We propose a bottom-up method to estimate the carbon footprint of digital agriculture scenarios considering deployment of devices over a diversity of farm sizes. It is applied to two use-cases and demonstrates that digital agriculture encompasses a diversity of devices with heterogeneous carbon footprints and that more complex devices yield higher footprints not always compensated by better performances or scaling gains. By emphasizing the necessity of considering the multiplicity of devices, and the territorial distribution of farm sizes when modelling digital agriculture deployments, this study highlights the need for further exploration of the first-order effects of digital technologies in agriculture. 
\end{abstract}

\section{Introduction}

The crossing of planetary boundaries~\cite{Richardson_2023} affects agriculture through yield loss~\cite{Ray_2019}, while agriculture contributes to their crossing. Limits such as biosphere integrity, or climate change are particularly concerned~\cite{Campbell_2017}.
Adapting agriculture to environmental challenges while reducing its negative impacts appears crucial for the sustainability of this sector.

Political institutions consider Information and Communication Technology (ICT) as a lever enhancing the efficiency, resilience, and sustainability of agriculture~\cite{muench_towards_2022}. The application of ICT to agriculture, digital agriculture~\cite{klerkx_review_2019}, encompasses technologies known as Smart Farming Technologies (SFT). According to~\cite{kernecker_experience_2020}, SFT notably includes GPS with real-time kinetics (RTK) guidance, variable rates inputs application, or autonomous robots.

Digital agriculture is expected to bring sustainability through process optimization, monitoring, and traceability. \citet{moysiadis_smart_2021} summarized related benefits as i) increased production, ii) reduced costs through input minimization iii) decreased labour efforts, and iv) enhanced final product quality. 

While SFT's adoption increases~\cite{nowak_precision_2021}, their ability to enhance sustainable food production remains debated due to a lack of concrete examples proving their added value to farmers~\cite{kernecker_experience_2020}. Digitalization alone may not suffice without integration into more sustainable production contexts, such as agroecological practices~\cite{schnebelin_how_2021,bellon_maurel_agriculture_2022}. Finally, SFT's benefits could also be balanced by the environmental footprint of ICT devices composing them.

Indeed, the ICT sector currently accounts for about 2 to 4\% of the global GHG emissions~\cite{freitag_real_2021} with a significant share coming from manufacturing. Few life cycle analyses (LCA) have been published regarding agricultural robots~\cite{pradel_comparative_2022,lagnelov_life_2021}. While they emphasize the need for a more consistent assessment of the footprint of digital agriculture devices, available data remain scarce and such studies do not offer a global view of the impacts associated with a widespread deployment of devices. Because such a massive deployment could be unsustainable~\cite{pirson_assessing_2021}, estimating the global consequences of a large-scale adoption of digital agriculture technologies is essential.

To address this gap, we propose a bottom-up methodology to estimate, within a similar agricultural context, the carbon footprint associated with the deployment of digital agriculture devices at a large-scale.

Our approach relies on a parametric inventory procedure (Section~\ref{method}) and a simplified impact assessment procedure accounting for embodied and usage emissions. It is primarily designed with the goal of comparing deployment scenarios with different technological complexities. This is demonstrated in Section~\ref{sectCasestudy} on two proof-of-concept examples at the scale of France, whose results are interpreted in Section \ref{Results}.

\section{Related works}
This section contextualizes our research paper and identifies the limits of existing works. It is divided into equipment surveys in field crop and dairy cattle farms, prospective scenarios, and existing methodologies.

\subsection{Equipment survey}

Regarding field crop farms,~\cite{nowak_precision_2021} studied the development of precision agriculture in the European Union (EU) and the United States and~\cite{schnebelin_usages_2022} examine digital tools used by 98 crop farms in the Occitanie region (France). Both studies show that global navigation satellite systems (GNSS), and management software are widely adopted by consulted farmers.

Regarding dairy cattle, the main used technologies are computer software (80\%), smartphone or tablet apps (60\%), GNSS (50\%), fixed and mobile sensors (including 40\% of cameras), and finally robots (30\%)~\cite{duroy_identification_2023}. Furthermore, 18\% of dairy producers use electronic tags, or radio frequency identification (RFID), and one-third of big farms (\textgreater 100 cows) use them.

\subsection{Prospective scenarios}
Few studies offer prospective scenarios for agricultural technologies. \citet{muench_towards_2022} present an innovation timeline expected by the EU between 2022 and 2050. This study relies on the concept of \textit{twin transition}, linking the possibility of a sustainable transition to a digital one. Likewise, in \cite{agrotic_5g_2021}, the authors relate to a ``deploy and see'' approach: by enabling growth of data flows, 5G technologies will reveal new usage for agriculture. 
Finally,~\citet{mora_foresight_2023} explore three different technological pathways able to permit a chemical pesticide-free agriculture in Europe by 2050.

Those studies commonly share a lack of details regarding the digital devices required to support the digital agriculture transition. \citet{muench_towards_2022} group emerging technologies by general categories (i.e. smart-farming, agroecological practices, AI-based solutions), without giving concrete details regarding the materials they might require. Similarly,~\citet{mora_foresight_2023} define their scenarios by general technologies, without further considerations. For instance, their \textit{global market} scenario implies autonomous robots acting on each plant, and their \textit{healthy microbiomes} one deploys a myriad of devices, sensors, and data. Those requirements still remain elusive and open to interpretations regarding the complexity associated with digital devices. 

By proposing a bottom-up method based on the materiality of digital agriculture devices, our study directly contributes to propose more concrete prospective scenarios, therefore overcoming identified limits of those works.

\subsection{Current approaches to assess ICT first-order effects} 

A general approach to assess ICT environmental footprint distinguishes first-, second- and third-order effects~\cite{hilty_relevance_2006, Berkhout_Hertin_2001}. Our study solely focuses on first-order effects, restrained to the carbon footprint, but other effects linked with studied use-cases are discussed in Section~\ref{sectDisc}.
First-order effects are associated with the life cycle of devices. They are always negative and can be assessed through attributional LCA (A-LCA) \cite{coroama_methodology_2020}. However, LCA-studies regarding SFT are scarce, limiting the possibility of estimating overall environmental consequences of a large deployment of those devices. Authors of \cite{pirson_assessing_2021} propose a bottom-up approach to overcome this limit. They model the footprint of the Internet of Things (IoT) edge devices according to their technological complexity. We use this concept of \textit{technological complexity} to compare devices on similar farming tasks. 

According to~\citet{Rasoldier_Combaz_Girault_Marquet_Quinton_2022}, models assessing sustainability should rely on scenarios, defined as reality simplifications based on relevant explanatory variables. Different scenarios can be explored provided that their differences and underlying hypotheses are explicitly stated. An application of that advice is presented by~\citet{guennebaud_assessing_2023}, comparing the power consumption of a fixed network infrastructure through a parametric model and different video-on-demand scenarios at the scale of a territory. The notions of scenarios and territory appear essential to explore and compare consequences associated to present and possible future technological deployments.

\section{Methodology}
\label{method}

The general goal of our method is to compare different deployment pathways of sets of digital devices, called technological systems (TS), over farms of a territory.
It is conceptually represented in~\figref{fig:conceptDiag}.
Embodied and use-related carbon footprints are estimated from a deployment simulation and a simplified life cycle inventory taking into account heterogeneous configurations.
The goal is not to provide a precise nor exhaustive assessment but rather to offer first-order footprint overviews enabling the exploration and quick comparison of many deployment variants.

We start by presenting a conceptual overview of our approach, before describing our inventory and assessment processes. We end this section with a high-level presentation on how scenarios are defined within our framework. Concrete examples are detailed in the next section.

\subsection{Conceptual modelling}
\label{conceptModel}

\begin{figure}[t]
 \centering
 \includegraphics[width=0.5\columnwidth]{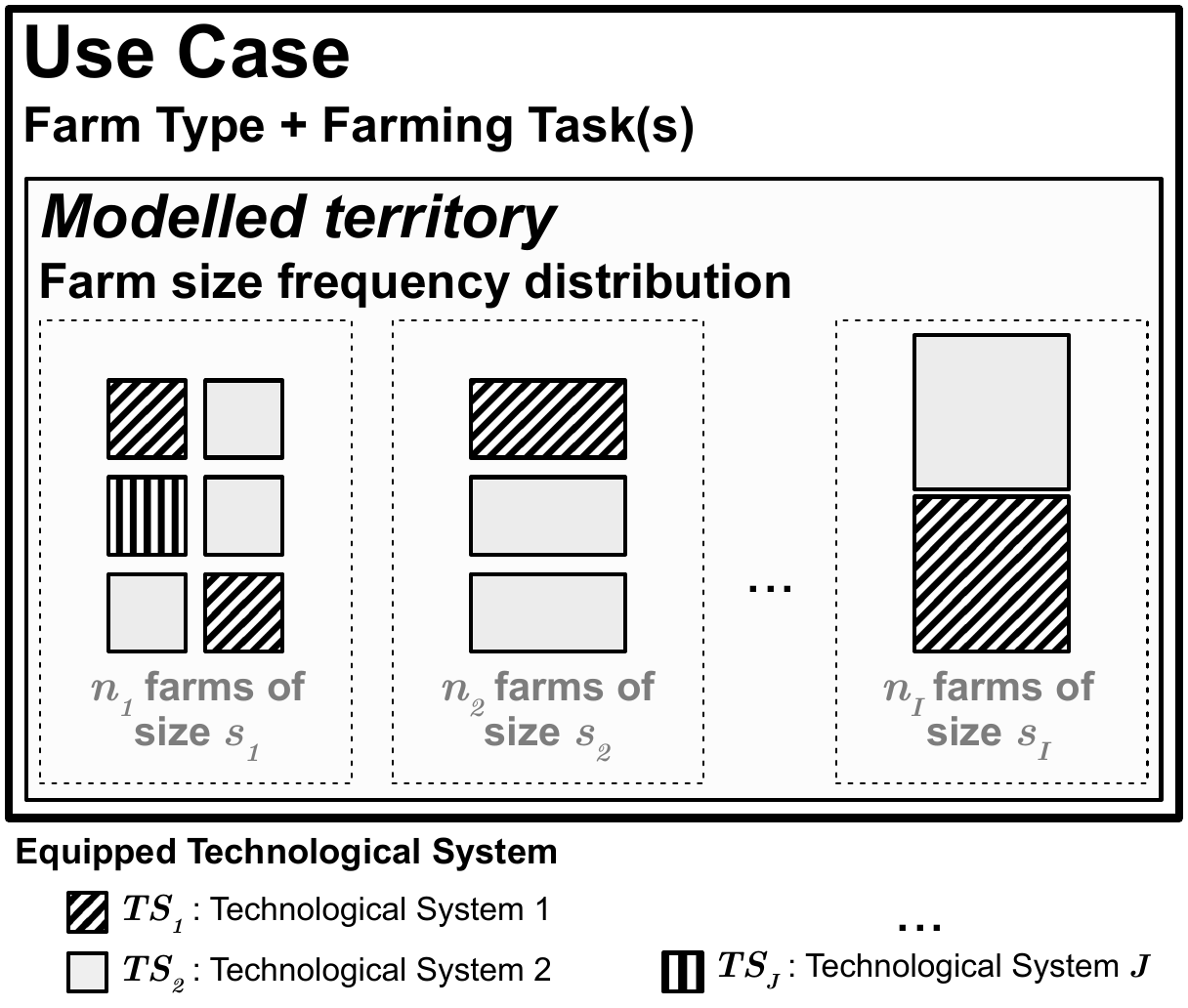}
 \caption{Illustration of the deployment of various TS over a distribution of farms of different sizes for a given use-case.
 }
 \label{fig:conceptDiag}
\end{figure}
Our modeling starts with the definition of
the studied \textit{use-case} by selecting a certain type of farm production and related farming tasks. Farming tasks are an entry point to identify technological systems. For instance, the dairy cattle use-case developed in Section~\ref{sectCasestudy} addresses individual cattle identification and heat detection tasks.

For selected farming tasks, we construct \TSj, the list of $J$ technological systems that we want to test in our scenarios.
A technological system is a collection of different digital devices (ICT, robots, IoT, autonomous systems...) deployed together on a given farm to perform a given farming task. Considering technological systems instead of isolated devices better reflects ICT structures and helps consider dependencies between devices.

In our approach, a territory is a set of farms of the use-case farm type. However, instead of simply working on a total number of farms with average statistics, we argue that it is important to explicitly consider the non-homogeneity in the farm characteristics because i) some TS might not be suitable for all farms, and ii) the deployment of a given TS within a given farm might also depend on the individual farm features. In this study, we distinguish farms of a given type solely based on their size, this parameter serving to scale a proper quantity of devices.
Following common practices~\cite{Eurostat_2022}, the size of animal-based production farms is defined per head and the size of plant-based farms is defined per hectares.
The set of farms is thus represented as a frequency distribution associating a discrete set of farm sizes $s_i$ ($i\in[1,I]$) with their respective quantity $n_i$ (\figref{fig:conceptDiag}). It serves as the input of our assessment pipeline, which mostly consists in instancing a device inventory list from which usage and embodied carbon footprints can be estimated based on the parameters of each device, as detailed in the next subsection.

\subsection{Inventory and Impact assessment procedures}
\label{assImp}

\begin{figure}[ht]
 \centering
 \includegraphics[width=0.9\textwidth]{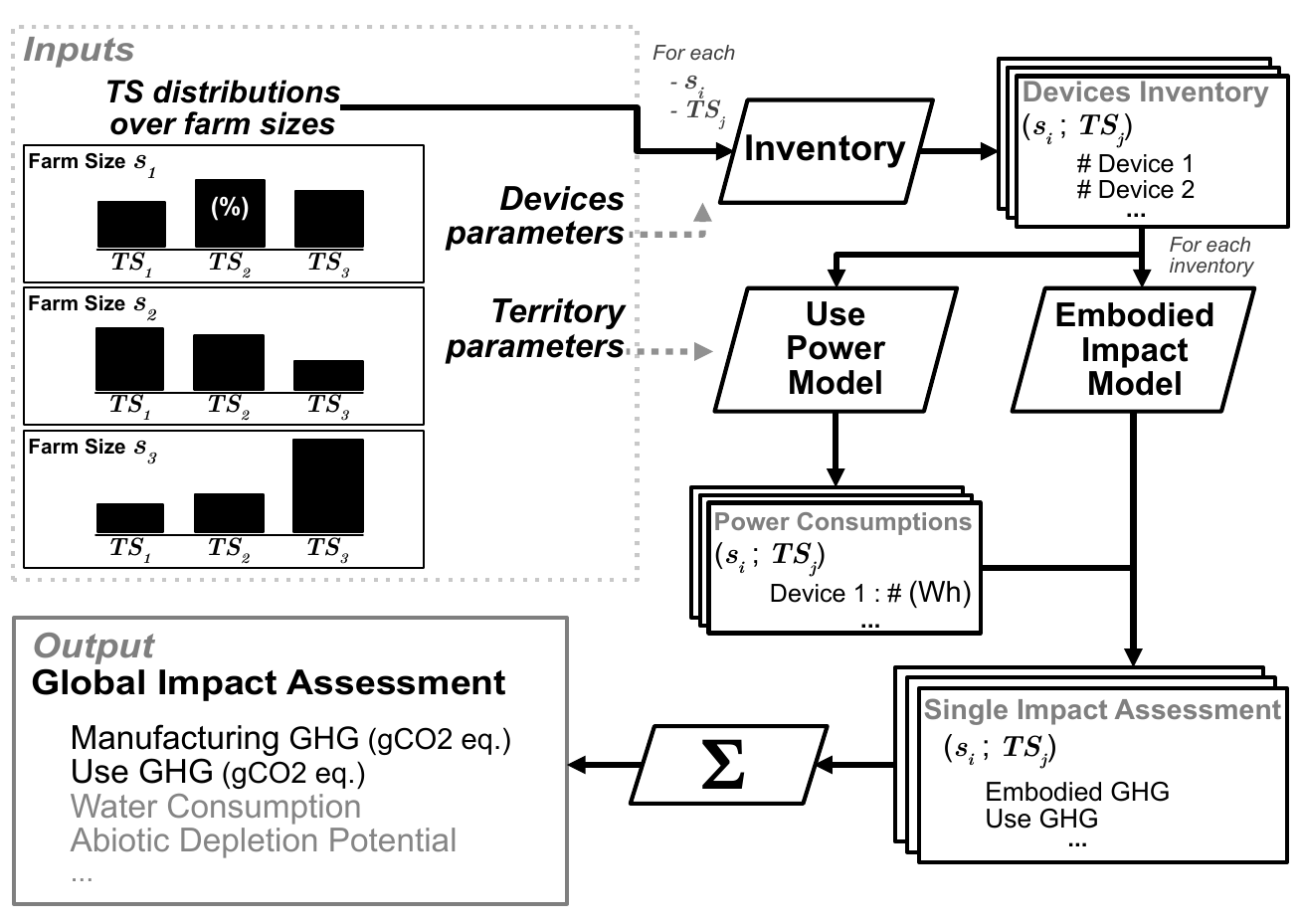}
 \caption{Schematized view of the process to assess the environmental impacts of a digital agriculture case study. \# represents a number. Inputs excluded, parallelograms indicate computing processes and rectangles indicate process outputs. Overlapping rectangles are used to represent a multiplicity of individual results.}
 \label{fig:methodprocess}
\end{figure}

\figref{fig:methodprocess} summarizes the process used to assess the carbon footprint of digital systems deployed over farms of a territory. 

\paragraph{Inventory}
The first processing step instances a device inventory, i.e. the total quantity of each type of device, depending on the TS, but also on the farm size.
For each subgroup composed of a farm size $s_i$ and a technological system \TSj, this process computes the quantity of each device required to equip a single farm of such a size with \TSj. This instancing mostly depends on the treatment capacity of each device. This capacity can be expressed in terms of the maximal number of cows supported by a given device (e.g. connected collar handle), or per hectare a day for crop treatments. These quantities are then scaled up by the number of farms of this subgroup which is obtained by the product: $n_i \times t_j(s_i)$, where $t_j(s_i)$ denotes the percentage of farms of size $s_i$ associated to \TSj. Those percentages can either directly come from real-world statistics of the adoption of current technological systems, or be manually refined to explore and compare future adoption pathways. We present such an editing tool in Section~\ref{sec:TSassociation}.
The inventory step is completed by use-phase power-consumption estimated through a dedicated \textit{power model}.

\paragraph{Impact Assessment}

Embodied GHG emissions are computed by dividing the related carbon footprint by the device lifetime (in years).
Annual energy consumption is converted to GHG emission based on the electricity carbon intensity of the territory. Annual carbon footprints from single impact assessments are summed up for global analysis.

\subsection{Mixed technological system modelling}
\label{sec:TSassociation}

When no statistics on TS deployments are available or when exploring hypothetical adoption pathways, the quantity $t_j(s_i)$ that represents the distribution percentages of \TSj over the farms of a given size $s_i$ must be manually defined. Setting each value individually would be very tedious, and nearly impossible when manipulating a dense set of farm sizes. We thus present here a simple tool allowing to quickly define them from a few parameters independent from the actual number of different farm sizes $s_i$.

\begin{figure}[t]
 \centering
 \includegraphics[width=0.7\columnwidth]{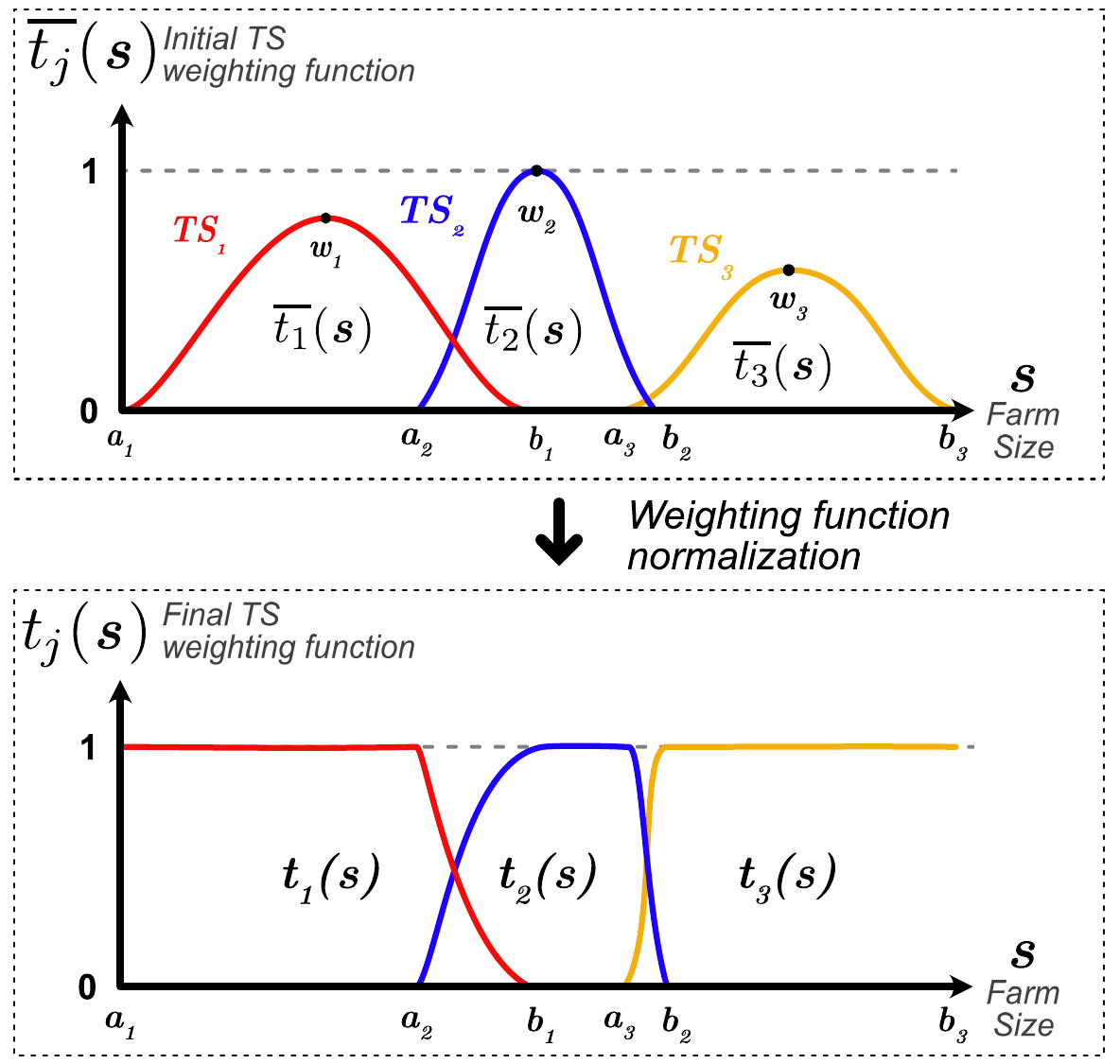}
 \caption{Schematized plots of mixed TS modelling with $J = 3$ TS. In this figure $s$ represents the set of all farm sizes, $\forall s_i, s_i \in s$, $a_j$, $b_j$, $w_j$ models initial TS weighting function
 $\overline{t_j}(s)$ and $t_j(s)$ TS final function after normalization.}
 
 \label{fig:formulas}
\end{figure}

In our current model, each farm is equipped with a unique TS among a list of $J$ technological systems. Hence, for a given farm size $s_i$, $t_j(s_i)$ must represent a probability mass function and we thus have to ensure that the percentages sum up to one, i.e. $\forall s_i, \sum_{j=1}^{J} t_j(s_i) = 1$.
To this end, we start by considering an intermediate weighting function $\bar{t}_j(s)$, that represents for any continuous value $s \in [s_1,s_I]$, the relative weight in $[0,1]$ of \TSj compared to the other TS. An example for three TS is depicted in~\figref{fig:formulas}-top. The mass function $t_j(s_i)$ is then obtained by normalizing the weighting function to sum up to one, i.e.: $$t_j(s_i) = \frac{\bar{t}_j(s_i)}{\sum_{l=1}^{J} \bar{t}_l(s_i)}.$$

As shown in~\figref{fig:formulas}, we chose for the $\bar{t}_j$ compactly supported bell-shaped functions controlled by three parameters only: a farm size range $[a_j, b_j]$ where the \TSj is present, and a weight $w_j \in [0,1]$ used to give higher or lower adoption weight to \TSj relatively to the others when multiple TS have overlapping farm size intervals.

For instance, in~\figref{fig:formulas}, \TS{2} is given a higher weight than the other two, implying sharper transitions in favour of \TS{2}.
Implementation-wise, we use for $\bar{t}_j$ the following quartic polynomial, but any other bell-shaped function would exhibit a similar behaviour:

\begin{equation}
 \label{quartfunction}
 \bar{t}_j(s) =
 \begin{dcases*}
 w_j \left( 1-m(s,a_j,b_j)^2  \right)^2 & if $s \in [a_j,b_j]$ \,,\\
 0 & otherwise\,,
 \end{dcases*}
\end{equation}
where $m(s,a_j,b_j)$ is a linear mapping of $s$ from $[a_j, b_j]$ to $[-1, 1]$.

\subsection{Exploring scenarios}
\label{ssscenarios}

The method presented so far assesses a territorial carbon footprint for a given use-case. By varying \TS{} and hypotheses through input parameters, it is possible to create and explore various scenarios and compare their carbon footprints.
More specifically, we can distinguish intrinsic versus extrinsic parameters.
The former category includes the composition of TS and the device's parameters such as powers, speed, or embodied impacts. Extrinsic parameters include the territory parameters such as electricity mix and farm size distributions, but also TS adoptions as modelled by our weighting functions. Farmer practices (e.g. number of control points for cattle identification, or pass frequency for crop fields) are also considered as extrinsic inputs.
The lifetime of equipment is ambiguous. It can be considered as intrinsic from the point of view of robustness, but also extrinsic if the actual lifetime is reduced through any kind of extrinsic obsolescence.

\section{Case Studies}
\label{sectCasestudy}

The goal of this section is to test the proposed methodology by comparing the deployment of digital agriculture devices exhibiting various \textit{technological complexity} levels, while fitting within a similar agricultural context.
Related TS were chosen either from existing or emerging technologies.
We tested two use-cases, dairy cattle and cereal crop farms, that are detailed below.
In both cases, system boundaries include on-farm deployed devices answering the identified farming tasks.

\subsection{Individual identification and heat metrics for dairy cattle}
\newcommand{\tabitem}{~~\llap{\textbullet}~~}
\begin{table}[ht]
\centering\renewcommand\cellalign{lc}
{
\small
\begin{tabular}{|p{2.5cm}|l|l|l|l|}
\hline
\multirow{5}*{\makecell{(a) TS FOR \\ DAIRY CATTLE \\ USE-CASE }} &
\textbf{TS\textsubscript{RFID}} & \textbf{TS\textsubscript{CC}} & \textbf{TS\textsubscript{JN}} & \textbf{TS\textsubscript{PC}} \\ \cline{2-5}
& Laptop & \multicolumn{2}{@{\hspace{1.5cm}}l|}{ Laptop (24/24)} & Laptop \\
\cline{2-5}
& \multicolumn{4}{@{\hspace{3cm}}l|}{ RFID chips + readers \textit{(ID)} } \\
\cline{2-5}
& &
\multirow{2}*{\makecell{ Connected collars \textit{(H)} \\ Collars antenna \textit{(H)} }} &
\multicolumn{2}{@{\hspace{1cm}}l|}{ Connected cameras \textit{(H)}} \\\cline{4-5}
&  &  & Jetson-Nano \textit{(H)} & PC with GPU \textit{(H)}  \\
\hline
\end{tabular}
\\
\vspace{0.1cm}
\hspace{0.01cm}
\begin{tabular}{|p{2.5cm}|p{2.78cm}|p{2.78cm}|p{2.78cm}|}
\hline
\multirow{2}*{ \makecell{(b) TS FOR \\ CROP USE-CASE }} &
\textbf{TS\textsubscript{BR}} & \textbf{TS\textsubscript{IR}} & \textbf{TS\textsubscript{AR}} \\ \cline{2-4}
& \makecell{ Base Robot} &
\makecell{ Intermediate Robot} &
\makecell{ Advanced Robot} \\
\hline
\end{tabular}
\vspace{0.1cm}
}
\caption{Details of modelled technological systems. (a) Device lists for the four TS used for the dairy cattle use-case. Considered farming tasks are Individual Identification~\textit{(ID)} and Heat Detection~\textit{(H)}. (b) Three tested TS for the crop use-case. Considered farming tasks are seeding and weeding.}
\label{tab:CaCrTS}
\end{table}

Our first use-case concerns TS related to individual cow identification and heat metrics for dairy cattle farms. Those technologies mainly help reduce work difficulty through process automation. We compare four different TS shown in~\tabref{tab:CaCrTS}(a).
The functional unit is: identify cows and collect heat metrics for the whole French dairy cattle farms during one year.

The first system, TS\textsubscript{RFID}, uses RFID tags to address individual identification. This system is included in all other TS of this use-case, due to its requirement for automation purposes. 
We assumed the use of passive RFID tags together with low-power readers.
It includes a laptop to store and manage the respective database. We account for 100\% of its impact, even though in practice it is expected to be used for other tasks, such as general farm management tasks (see also Section~\ref{sectDisc} (§3)).
The second one, TS\textsubscript{CC}, adds the use of connected collars (CC) to address the heat metric function\footnote{Those collars and their required infrastructure were inspired from a field study and by industrial solutions like~\cite{AllflexSenseHub}.}.
The last two systems use cameras and computer vision (CV) algorithms to tackle both functions. RFIDs are kept for superior robustness. TS\textsubscript{JN} uses low power Jetson-Nano boards for cattle recognition, with one board per camera.
TS\textsubscript{PC} replaces the multiple Jetson-Nano boards with a single PC equipped with GPU scaled depending on the number of cameras deployed~\cite{Qiao_Su_Kong_Sukkarieh_Lomax_Clark_2019}. A laptop is still used for general purposes but with less use per day and an extended lifespan. This more power-intensive TS variant is expected to provide more robustness and opens the door to other monitoring functions as studied by recent academic papers~\cite{Dac_GonzalezViejo_Lipovetzky_Tongson_Dunshea_Fuentes_2022, Pasupa_Lodkaew_2019}. 

\subsection{Seeding and weeding Robots for field crop farming}

This use-case models robots used for seeding and weeding in field crop farms. Agricultural robots are an emerging technology expected to reduce workload, soil compaction, and input consumption (e.g. fertilizer, oil, pesticides) through automation and improved precision. As presented in~\tabref{tab:CaCrTS}(b), we designed a series of three robots in an incremental way along a ``technological complexity'' axis. They are all based on the same power source (i.e., electricity) and base-platform, but differs in capabilities through technological features. We approximate similar task efficiency between robots, based on the results of \citet{Gerhards_Risser_Spaeth_Saile_Peteinatos_2024}.
The functional unit is: seeding and weeding crop fields for the whole French cereal crop farms during a year.

The \textit{Base Robot} (\TS{BR}) is inspired by the~\citet{Farmdroid_2023} and models a robot using relatively common technologies (i.e., GNSS and RTK-correction). Its low power leads to slower treatment capacity but extended autonomy.
The \textit{Intermediate Robot} (\TS{IR}) is inspired by the \textit{Orio} from~\citet{Orio_2023}. The use of two LiDAR sensors enables the robot to go faster while remaining safe.
The \textit{Advanced Robot} (\TS{AR}) is inspired by the \textit{RobotOne} of~\citet{PixelfarmingRobotics_2022}. This version extends the previous one with \textit{smart} tools, enabling more precise and selective weeding. We assume such a technology could enable reducing frequency passes and power requirements, at the expense of a slower pace, and hence a lower daily working capacity.
Note that mentioned existing models are only inspiration sources for the modelling of fictional robots incrementally built.

\subsection{Inventory, Assessment and Modelling Assumptions and Data}

\paragraph{Life Cycle Inventory and Modelling Assumptions}
Both use-cases consider France as the modelled territory for farm size distributions.
The dairy cattle use-case relies on the farm size distribution given by the French Ministry of Agriculture~\cite{Depeyrot_Perrot_2020} for a total of 51,000 farms and roughly 3,458,000 heads. The crop use-case uses cereal farm size distribution from~\citet{Agreste_2016}. In this latter case, we omit small farms (\textless 20 ha) assuming that most of them are not concerned by the studied robotic systems, as robots are currently expensive and designed to treat larger surfaces. To be less sensitive to arbitrary thresholds from \cite{Agreste_2016}, we reconstructed a $C^0$ dense distribution for every $s_i$ in $ \left[20,400\right] $ (in ha) reproducing the given coarse statistics with a total of 65,223 farms and 7,358,412 hectares.

Modelled TS are inspired by equipment surveys and existing product fact sheets related to cutting-edge technologies. Associated values were estimated from technical documentation,
and academic LCI papers \cite{Lagnelov_2023}.

For the cattle use-case, we assume the RFID system is active 11.5 hours a day and stays in sleep mode the rest of the time. Heat-detection systems and related ICT devices, however, are assumed to be on 24h a day for monitoring purposes.

For the crop use-case, we model robots from their electronic components (sensors, cameras, embedded computing units, etc.). We consider them equipped with the same solar panel surface. \TS{MR} and \TS{AR} batteries are scaled to ensure 10 hours of daily autonomy. Due to its much lower power requirement and consumption (very slow speed), the battery of \TS{BR} enables 16 hours of autonomy. The technological complexity of those systems differs regarding the number of sensors, processing capacities, and engines. Considered power affects weight and achievable speed, positively affecting efficiency. To ensure consistent hypotheses regarding parameter relations when no data is available, such as how power affects autonomy, proportionality rules are applied.

Whereas ICT devices (e.g. laptops or PCs) are deployed independently of the farm size, most equipment quantities are scaled based on a maximal capacity expressed in the number of heads for the cattle use-case, or in hectares per hour per device for the crop use-case. Some device quantities are scaled relatively to the number of other devices (e.g. GPU and Jetson-Nano boards depend on the number of cameras). Device parameters and further modelling assumptions are detailed in Supporting Information S1.

We denote as $q_d(s_i)$ the amount of a given non-robotic device $d$ per farm of size $s_i$. The annual power consumption of this quantity of devices is estimated as $$q_d(s_i) \times 365 \times \left( Tactive_d \times Pactive_d + Tsleep_d \times Psleep_d \right) \,,$$
using active (resp. sleep) powers $Pactive_d$ (resp. $Psleep_d$) and active (resp. sleep) hours per days $Tactive_d$ (resp. $Tsleep_d$).

Similarly, we denote $q_r(s_i)$ the required quantity of a given robotic device $r$, for a farm of size $s_i$. $q_r(s_i)$ is estimated from the hourly treatment capacity $C_r$ (in hectares per hour per device), the maximal working time per day $Tactive_r$ (in hours per day), and a use periodicity at seasonal peak time $U_r$ (in days):
$$q_r(s_i) = \ceil[\bigg]{\frac{s_i}{U_r C_r Tactive_r}} \,.$$
The yearly power consumption of a robotic device $E_r$ is estimated using the number of passes $D_r$ per year on the same area, the potential solar panel power supplement $Esp_r$ (in Wh/day), and distinguishing active (work) versus travel (i.e. without work) times:

\newcommand{\totalworktime}{{Ttotal}_{r,i}}
$$E_r(s_i) = D_r q_r(s_i) \left(
  Pactive_r \totalworktime
 + (Ptravel_r \times Ttravel_r - Esp_r)\times \ceil[\bigg]{\frac{\totalworktime}{Tactive_r}}
\right) \,,$$
where $\totalworktime = \frac{s_i}{q_r(s_i) C_r}$ is the total work time (in hours) per robot to treat the whole surface $s_i$. The fractional part represents the number of days to treat the whole surface. We supposed a plough-free biological agricultural context to establish the number of passes $D_r$ = 9, based on~\cite{ChambresNormandie_2024}. Each pass has to fit the number of days at seasonal peak time $U_r$ based on the robot capacities. As detailed in Supporting Information S1, $Tactive_r$, $Ttravel_r$ and $C_r$ are computed from lower-level parameters such as treatment width, speed and distances.

\paragraph{Life Cycle Impact Assessment}
For the use phase, we considered the French electricity mix\footnote{We use $68$ gCO2e/kWh computed as the average over the 2017-2022 period \cite{electricitymap}.}. 
Best efforts were put in finding embodied data from public LCA databases\footnote{Base Empreinte \cite{ademeBaseEmpreinte}}, and academic LCA papers \cite{Bottani_Manfredi_Vignali_Volpi_2014,Lagnelov_2023}. Despite coming from different databases, chosen data consider similar life cycle stages. Finally, as we aim to propose first carbon footprint magnitude orders, we use weight-based public average carbon impact factors to compute our results.

\section{Results}
\label{Results}

This section presents and interprets results obtained from applying our method to the two selected use-cases.

\subsection{Analysis of full deployment scenarios}
\begin{figure}[ht]
 \centering

 \includegraphics[width=\columnwidth]{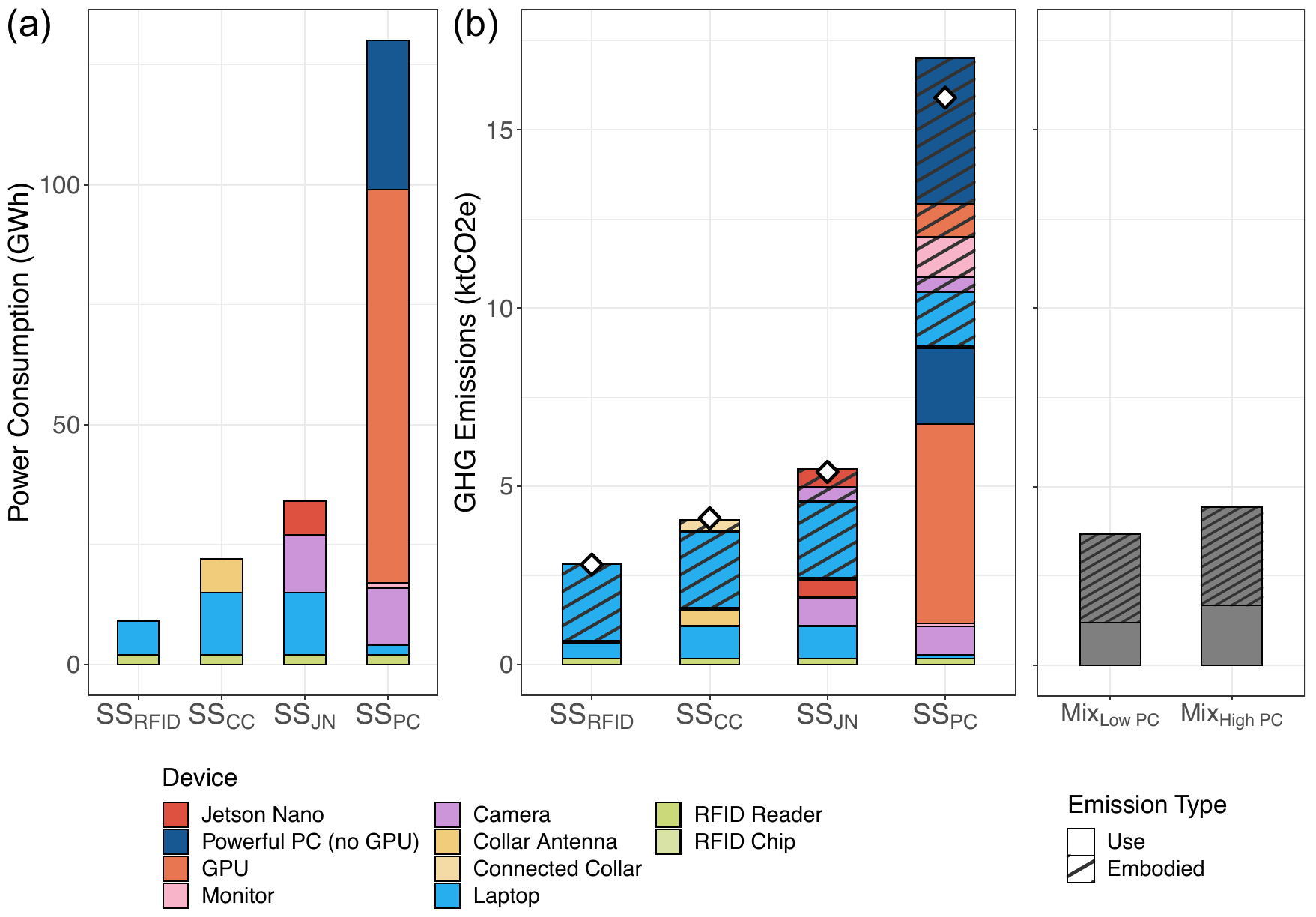}

 \caption{Power consumption (a) and GHG emissions (b) for cattle use-case using French dairy cattle farm distribution (51,000 farms) and electrical carbon footprint for one year. (a) and (b) left-column: Full deployment of each technological system using farm size distribution. ($\diamond$): GHG emissions using an average farm size extrapolation approach. (b) right-column: Two different scenarios mixing the considered technological systems. "Low PC" mix deploys more systems using Connected Collars (\TS{CC}), "High PC" mix deploys more systems using PC for computer vision (\TS{PC}). Details related to mixed scenarios are available in Supporting Information S2 and figure's data is available in Supporting Information S3.}
 \label{fig:catfavg}
\end{figure}

\begin{figure}[ht]
 \centering
 \includegraphics[width=\columnwidth]{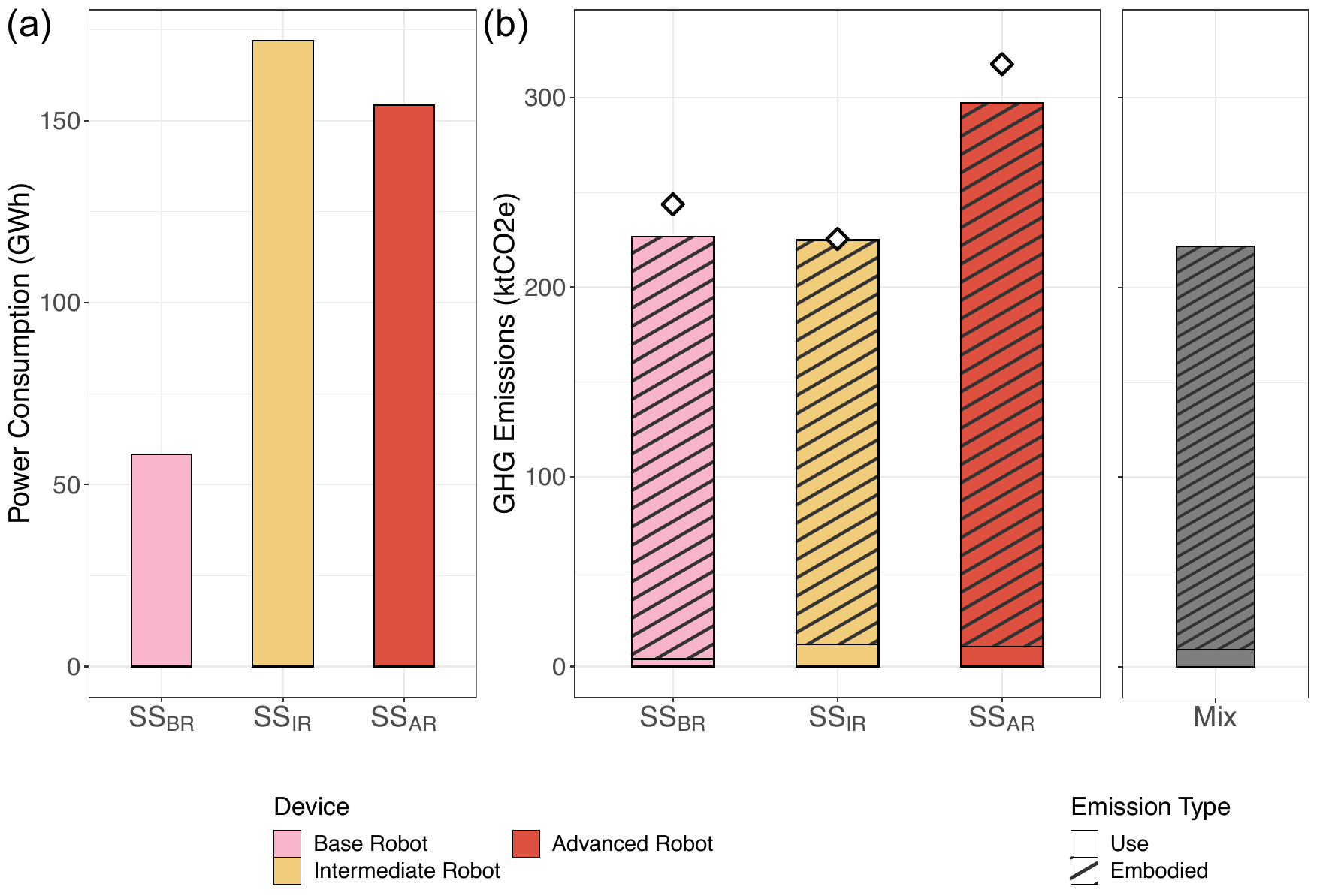}
 \caption{Power consumption (a) and GHG emissions (b) for crop use-case using truncated French cereal farm distribution (65,223 farms) and electrical carbon footprint for one year. (a) and (b) left-column: Full deployment of each technological system using farm size distribution. ($\diamond$): GHG emissions using an average farm size extrapolation approach. (b) right-column: A scenario mixing the three considered technological systems. Details related to mixed scenario are available in Supporting Information S2 and figure's data is available in Supporting Information S3.}
 \label{fig:cropfavg}
\end{figure}

Panels (a) and left-column of (b) of Figures~\ref{fig:catfavg} and \ref{fig:cropfavg} compare the power consumption and the carbon footprint of deploying a unique TS over all farms. Right-column of (b) considers mixed scenarios. Details about mixed scenario distributions are available in Supporting Information S2. Single TS scenarios are prefixed by ``\Ss{}''. Both figures report breakdowns at the device level.

For the cattle use-case, an expected observation is that GHG emissions increase proportionally to technological complexity. Left-column of panel (b) reveals that for single TS scenarios, GHG emissions are dominated by the manufacturing phase of the laptops, except for \Ss{PC} for which GPU power consumption significantly increases the use-phase GHG emissions, despite a low carbon electricity mix. While RFID chips are the most deployed devices, their impact appears marginal compared to other devices. This can be due to their relative technological simplicity. Use-phase-related GHG emissions are linearly linked to the use-phase electrical power consumption. Most emitting devices are still ICT devices, but also always-on equipment such as cameras and antennas. 

The crop use-case exhibits similar behaviours. It offers supplementary insights regarding the influence of technological complexity over large-scale carbon footprint. 
The first panel shows that power consumption globally increases with technological complexity and achievable speed, \TS{IR} being the most power-consuming robot. However, in a context of low-carbon electricity mix, and because robotic devices have high embodied carbon footprints, total GHG emissions (panel (b)) are more linked with the number of deployed devices, and the embodied footprint, than the power consumption (panel (a)). In this context, \Ss{IR} appears to be the least impactful system. Its higher capacity enables the deployment of fewer devices. The lower use impact of \Ss{AR} and its extended work periodicity cannot compensate for its lower range and higher embodied footprint. This leads to an increase of its total carbon footprint. 

Comparing these two use-cases, the crop use-case (\figref{fig:cropfavg}) exhibits higher orders of emissions than the cattle use-case.
This is not only explained by the higher number of farms ($14,000$ more), but also by the choice of TS that implies heavier and more consuming devices.
Thus, behind the same concept of \textit{digital agriculture} comes very different devices with GHG emissions of heterogeneous orders of magnitude.
This observation illustrates the need to distinguish use-cases and technological systems studied when assessing the carbon footprint of digital agriculture.
Nevertheless, the narrow scope of these two studies does not allow making any further comparisons between them.

\subsection{Distribution-based versus average-based approaches}
The panels (b) of Figures~\ref{fig:catfavg} and \ref{fig:cropfavg} also include a comparison to another assessment method, based on an average extrapolation. It estimates the GHG emissions of an average farm size and extrapolates the global emissions based on the number of farms. These extrapolations are shown with white diamonds ($\diamond$).
Whereas the average-based method seems to work reasonably well for the first use-case (except for \Ss{PC}), it breaks for the second one with an overestimation (for \Ss{BR} and \Ss{AR}) depending on the deployed TS. 
This inconsistency justifies the need to explicitly account for farm size distribution even if the purpose is to make relative comparisons only. It calls to further study the impacts of the evolution of farm distributions over a territory.

An explicit farm size distribution further allows us to test more realistic scenarios mixing multiple TS over farms, using the weighting mechanism explained in Section~\ref{sec:TSassociation}.
For the cattle use-case, we designed two such mixed scenarios. Their TS distributions over farm sizes are available in Supporting Information S2.
Both scenarios assume that more advanced monitoring systems are mostly required by the largest farms.
The main difference between the ''Low PC'' and ''High PC'' scenarios is that the former assumes a slightly higher adoption rate of the collar-based TS, but a marginal adoption of the PC-based TS.
'Low PC'' thus relies on a lower technological level than ''High PC''. Right-column of Figure~\ref{fig:catfavg}(b) shows that both scenarios exhibit lower GHG emissions than the singular \Ss{PC} scenario, but ''High PC'' comes with a substantially higher footprint than ''Low PC''.

For the crop use-case, we designed a single mixed scenario, shown in~\ref{fig:cropfavg}-(b) right-column, minimizing GHG emissions. This scenario distributes systems with Base Robot (\TS{BR}) for farms under 120~ha, and Intermediate Robot (\TS{IR}) for larger farms. Under chosen hypotheses, \TS{AR} is therefore never more advantageous than \TS{BR} or \TS{IR} regardless of the farm size. Consequently, its reduced frequency passes and power requirements cannot compensate for its slower pace and higher embodied footprint.

\subsection{Cattle use-case efficiency}
\begin{figure}[ht]
 \centering
 \includegraphics[width=0.8\columnwidth]{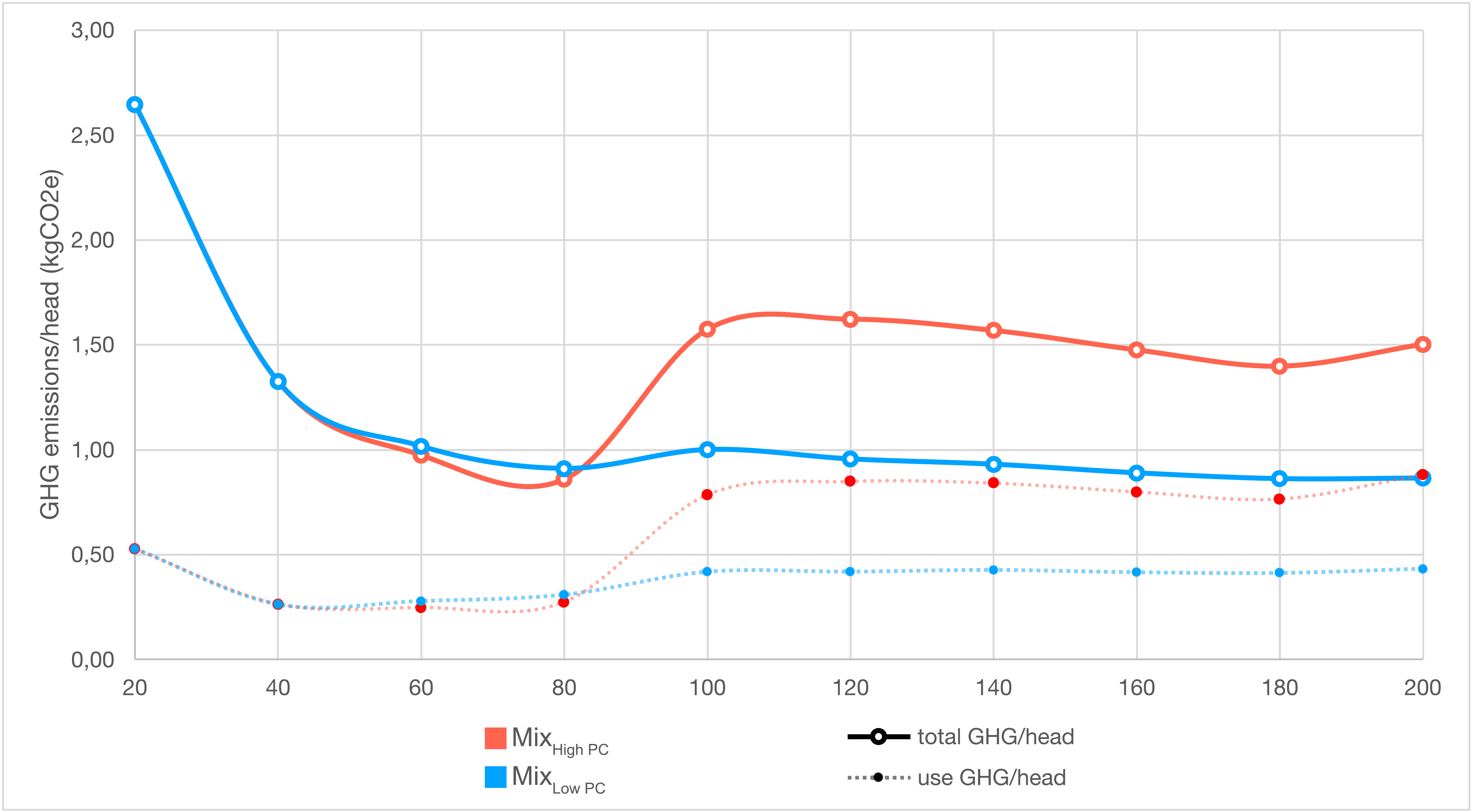}
 \caption{GHG emissions per head for cattle use-case comparing ''Low PC'' and ''High PC'' mixed scenarios. Use GHG emissions are displayed in dashed lines. Figure's data is available in Supporting Information S3.}
 \label{fig:cattleeff}
\end{figure}

To get insights on the relative environmental efficiency of the ''Low PC'' and ''High PC'' mixed scenarios with respect to farm sizes, we computed in Figure \ref{fig:cattleeff} the GHG emissions per head for each category of farm size.
We observe a rapid drop of the GHG emissions per head when moving from very small to medium farm sizes. This is primarily due to the GHG emissions of the laptops which are amortized on a larger number of heads. However, this apparent drop is more likely the result of a limitation of our model where 100\% of the laptop impacts are allocated to the dairy cow management. This might be acceptable for large enough farms dedicated to milk production, but the smallest farms with about 20 heads only are expected to carry on other productions (e.g. vegetables or cheese) for which this laptop is also required for general management tasks, meaning that for such small farms it is unfair to allocate 100\% of the laptop impacts.

For farms of 60 heads or more, the efficiency either stagnates because of the expected need for more sophisticated and impactful technologies, or even decreases when the heavy PC-based TS starts to be adopted by a non-negligible share of farms (in this case, about one-third).
As shown by the dotted lines, this rebound is even more pronounced for the power consumption per head.

This experiment shows that the well-known concept of \textit{scaling gains} might not always hold, especially if scaling calls for more technologies to keep the system manageable.

\subsection{Sensitivity analysis for the crop use-case}

\begin{figure}[ht]
 \centering
 \includegraphics[width=0.8\columnwidth]{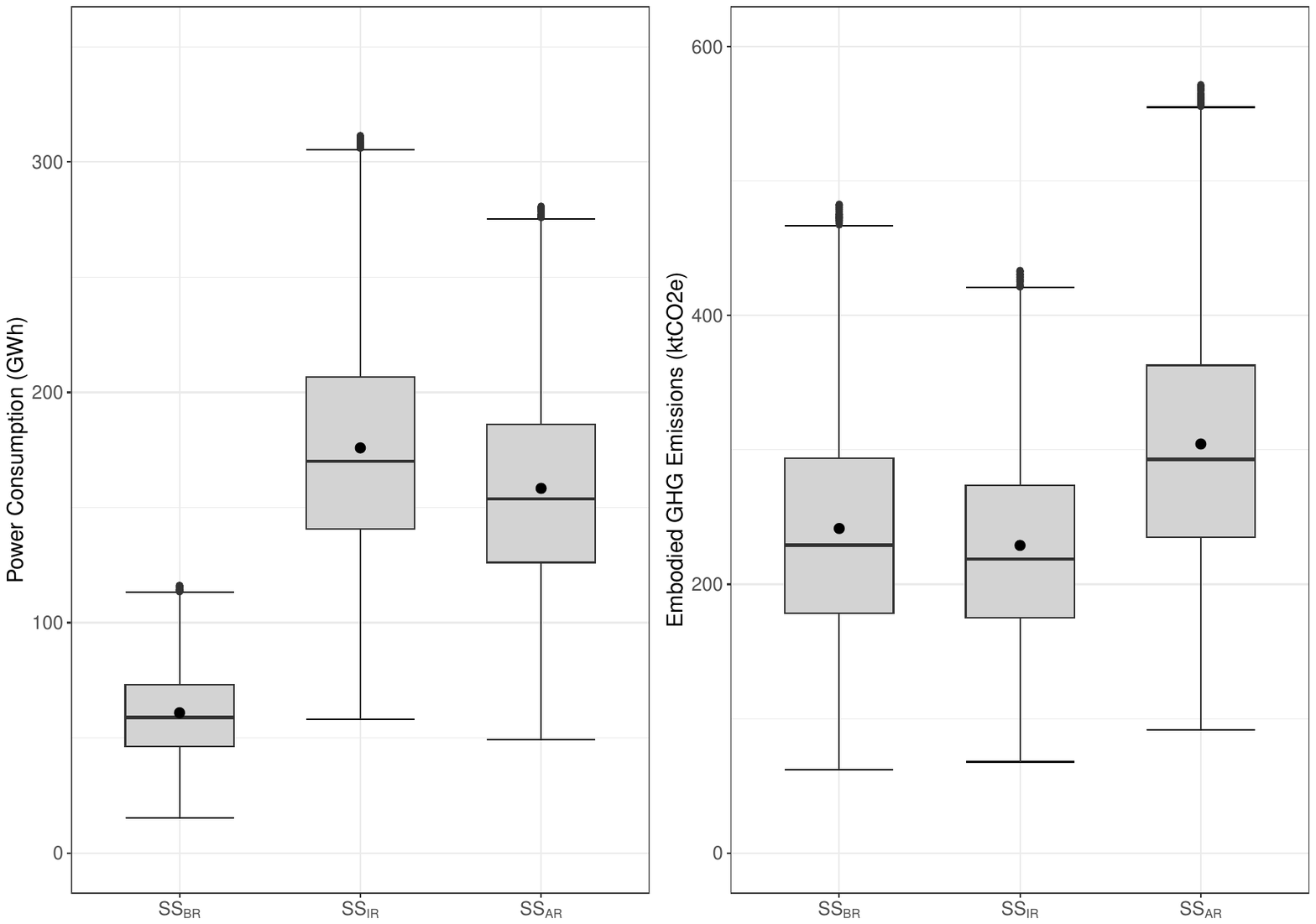}
 \caption{Power Consumption and embodied GHG emissions sensitivity analysis for the crop use-case using French farm distribution and electrical carbon footprint for a full deployment of each system using a Monte Carlo algorithm (n=10,000). Variations randomly affect lifespan, power consumption, photovoltaic production, use periodicity and device capacity with a relative standard deviation of 20\%. Figure's data is available in Supporting Information S3.}
 \label{fig:crophyp}
\end{figure}

Results of our model highly depend on input parameters and hypotheses. To get some insights on the variability, we conducted a Monte~Carlo-based sensitivity analysis (with 10,000 random samples). \figref{fig:crophyp} presents results for the crop use-case. Results for the cattle use-case are available in Supporting Information S2. Variations are made on device capacity, lifespan, work power consumption, and solar panels production, log-normal distributions with a relative standard deviation of 20\%. Work periodicity is also tested with random variations of more or less one-day intervals.
\figref{fig:crophyp} shows significant variations depending on the set of values. 
Whereas \textit{BaseRobot} exhibits a clear lower power consumption, there is no clear hierarchy regarding embodied emissions.
This is mostly explained by the influence of treatment capacities over the amount of required devices, especially for the base (low emissions per unit, but small labour capacity per unit) and advanced robots. Embodied emissions are also sensitive to lifespan variation.
Depending on the technical specification of a device, one category might appear less impactful than the others and \textit{vice versa}. The resulting uncertainty calls for better access to technical specifications and capacities to improve impact assessment related to deploying digital agriculture systems.

A complementary sensitivity analysis about changes in the electricity mix for the use-phase emissions of the crop use-case can be found in Supporting Information S2. It compares the electricity mix of four European countries with significant agricultural production (France, Spain, Germany and Poland). Electricity mixes were computed as the average over the 2017-2022 period \cite{electricitymap}. The operational consumption grows positively depending on the chosen electricity mix and remains lower than the fixed embodied carbon footprint.

Even if our results are partial and should be considered as gross estimations, they show that thinking TS deployment in terms of distributions over different farm sizes is a relevant approach to study scaling effects. They highlight the need to consider digital agriculture devices in their diversity of size and complexity, as this leads to heterogeneous carbon footprints. Finally, this work calls for better access to data and confirms the need for a better understanding of the large-scale consequences of digital agriculture.

\section{Discussions}
\label{sectDisc}

We proposed a methodology to estimate the direct carbon footprint related to the deployment of digital agriculture over a territory. Our findings indicate that considering farm distributions instead of average extrapolations holds more precise results. In the context of low data availability, our bottom-up approach offers overall estimations for a certain territorial area. We applied this methodology for two use-cases in France to propose a first-of-its-kind carbon footprint estimation of digital agriculture systems. Each quantification is given assuming comparable system efficiency and agricultural contexts. While those preliminary results are encouraging, lots of improvements can be made. 

\paragraph{Beyond carbon footprint}
This study is restrained to an overall estimation of GHG emissions produced by the manufacturing and the use of digital technologies deployed over a territory. However, when studying environmental impacts of digital devices, other environmental criteria such as water consumption, eco-toxicity or abiotic resource depletion appear to be at least of equal concerns. Including them in our model would be particularly interesting as the dynamic of the results could be very different for such indicators. Those indicators could notably allow us to quantitatively compare benefits and pitfalls linked to the deployment of digital technologies for agriculture. However, this addition would require more publicly available data regarding the manufacture and use of considered technologies.

More generally, social and economic impacts such as cost and dependencies, as well as resilience or autonomy criteria could help to better seize the impacts of digital agriculture at large-scale.

Hence, we emphasize that despite providing some rough orders of magnitude for simplified and narrowed use-cases, our preliminary results on power consumption and GHG emissions do not reflect the large-spectrum of consequences linked to deploying digital technologies at a large-scale.

\paragraph{Beyond farm size distribution}
Modelled agricultural activities were based on the distribution of related farm types. While this approach enabled more precise results than average extrapolations, our model could gain further precision by considering a wider range of farm productions and more complex device management such as pooling. This would help us to better consider agricultural activities at a territorial scale. A huge improvement could be considering changes in agricultural practices, hence enabling the implementation of scenarios modelling non-digital solutions.

\paragraph{Beyond technological systems}
Our model needs to consider a wider range of TS and farming tasks. While farming tasks properly worked for the cattle use-case, due to mono-functional devices, the crop use-case leads us to go beyond this approach, as considering all farming tasks were done by each system (weeding and seeding). Additionally, farming task approach appears limited by its narrowness, as it leads to allocation issues. One example is the shared laptops of the cattle use-case. As our goal is to give absolute magnitude orders of prospective deployment, a better approach would be to extend the study boundaries to include all relevant shared equipment. In our context, this means to include all activities present on farms.

To better understand the impacts of digital agriculture, it will also be necessary to consider networks and data-centre activities in addition to on-farm activities. 

\paragraph{Variability and uncertainties}
As observed in the result section, considered digital devices present high heterogeneity of carbon footprints. In line with~\cite{pirson_assessing_2021}, digital agriculture devices should therefore not be seen as a homogeneous whole, even for similar tasks, as their impacts seem to be linked with their technological complexities.

However, even when looking at a single technological system, many sources of variability and uncertainties accumulate.
Firstly, models offer a simplified vision of a much more complex reality, and ours is not an exception.
Secondly, the lack of precise technical specification about the composition of digital agriculture devices leads to assumptions introducing a strong variability bias.
Thirdly, properly assessing the environmental impacts of each of these heterogeneous components would require a tremendous amount of work, and rough approximations have to be made.
Fourthly, the actual capacities, farming practices, and lifespans play an important role on the results while being challenging to precisely capture, as our sensitivity analysis shows. 

Upon to that, territorial factors such as farm size distributions, \TS{} deployment strategies, or carbon intensity of the electricity grid may affect the global assessment. Hence, our observations could be different when considering another country than France. 

\paragraph{Putting our results into perspective}

A more thorough assessment of the impacts of a TS requires widening the range of considered effects and contextualizing them within the broader sector. Considering our results for the crop use-case, first-order carbon footprint (i.e. including manufacturing and use phases) of globally deploying modelled robots varies between $0.2$ and $0.3$ Mt~\COtwoe{} per year.
In comparison, the average annual GHG emissions resulting from the combustion of machinery, engines and boilers utilized across the entire French agricultural sector, between 2014 and 2022, amount to 9.8 Mt~\COtwoe~\cite{Citepa_2023} per year. By narrowing down this global scope to the surface considered in this paper for the crop use-case (around $7.3$ million hectares) and focusing solely on the kind and number of passes defined in our study for seeding and weeding farming tasks, the use of thermal tractors and associated mechanical equipment for treating this surface would yield approximately 
$0.92$~Mt~\COtwoe over one year. This estimate is based solely on typical fuel consumption (Non-Road Diesel or NRD) per hectare for different crop field operations~\cite{Chambresdagriculture_2022}.
Additional details are available in Supporting Information S2. 
Hence, only considering GHG emissions of a restrained use-case, the manufacturing and use of electric robots to substitute thermal engines for certain tasks might significantly reduce emissions by diminishing NRD consumption. In addition of input reduction and task automation, this substitution of heavy thermal vehicles by lighter ones could reduce soil compaction and associated emissions~\cite{Pulido-Moncada_Petersen_Munkholm_2022}.

However, significant power disparities exist between electric and thermal vehicles, making a direct substitution of thermal tractors by electrical robots impractical according to~\cite{Scolaro_2021}. While studied robots may handle seeding, weeding, or narrow soil work with satisfactory results, they cannot replace powerful thermal engines for large-scale demanding tasks or deep soil works. Something which was not modelled in this study while being important regarding limits about the tractor-robot substitution is the need for tractors to transport robots, considered as non-road vehicles, between separate fields or from the farmhouse to the plots as studied by~\cite{pradel_comparative_2022}.
Existing uncertainties regarding this substitution could lead to robots stacking over existing devices, justifying the comparison of NRD consumption to the robots' entire life-cycle. The comparison could also give different results considering other indicators than GHG emissions.
Lastly, considering the economical investment required for robotic devices, the deployment of such technologies could accelerate the French contemporary trend, concentrating agricultural production around fewer farms of larger area~\cite{Le_Guern_2020, kernecker_experience_2020}, while implementing more sustainable agricultural practices appears easier on small-sized farms according to~\cite{DeSchutter_2014}.

\section{Conclusion}
We presented a methodology to estimate the carbon footprint of digital agriculture systems at the scale of a territory, hence better incorporating the diversity of farm sizes and technological system adoptions.
Our results confirmed that this approach yields to more precise and realistic assessments than a simpler average-based method. They reveal a significant variability of GHG emissions depending on the technological complexity of devices and their respective capacities.
In the case of dairy-cattle monitoring, we showed that more advanced devices exhibit higher footprints, and, provided that larger farms call for more advanced equipment, this higher footprint is not compensated through scaling gains offered by large farms.
Presented use-cases are proof of concepts of our approach, and this paper discussed several opportunities to improve this current version by extending its scope both in terms of system boundaries and impact indicators.
Our methodology could provide better insights to decision makers regarding global consequences associated with the digitalization of agriculture, therefore enabling them to take more informed decisions toward agriculture sustainability.

\section*{Conflict of Interest}
The authors declare no conflict of interest.

\section*{Supporting Information}
Supporting information is linked to this article on the JIE website:

\begin{framed}
\textbf{Supporting Information S1:} This supporting information details the device parameters used for the scenarios of this study.

\textbf{Supporting Information S2:} This supporting information provides mixed scenario distributions, complementary sensitivity analysis and additional discussion results.

\textbf{Supporting Information S3:} This supporting information provides data used for figures 4, 5, 6 and 7.
\end{framed}

\section*{Data Availability Statement}
All data used in this study are publicly available and freely accessible from the sources cited.

\bibliographystyle{unsrtnat}
\bibliography{article}

\end{document}